\documentclass[aps,numerical,graphicx,showpacs,preprint,prl]{revtex4-1}
\usepackage{amsmath}
\usepackage{graphicx}
\usepackage{hyperref}

\begin{document}

\def\bib{B\kern-.05em{I}\kern-.025em{B}\kern-.08em}
\def\btex{B\kern-.05em{I}\kern-.025em{B}\kern-.08em\TeX}
%

\title{
Fundamental quantum and relativistic formulation of thermal noise and linear conductance in an 1D quasi-particle ensemble under ballistic transport-regime
}
\author{Lino Reggiani}
\affiliation{Dipartimento di Matematica e
Fisica, "Ennio de Giorgi" \\ Universit\`a del Salento, via Monteroni
73100 Lecce, Italy.
}
%
\author{Federico Intini}
\affiliation{Department of Engineering, "Enzo Ferrari”,
University of Modena and Reggio Emilia,
Via~P.~Vivarelli, 10, 41125 Modena, Italy
}
\author{Luca Varani}
\affiliation{IES, Univ. Montpellier,  \\
860 rue de St Priest, 34095 Montpellier cedex 5, France
}
\begin{abstract}
We investigate quantum and quantum-relativistic effects associated with the noise power spectrum and the fluctuation--dissipation relation between current--noise spectra and linear--response conductance at low frequencies of the electromagnetic field.
At high frequencies, vacuum catastrophe is shown to be avoided by the presence of Casimir force.  At low frequencies, the quantum effect associated with one--dimensional structures under the conditions of ballistic transport typical at the nanometric scale length are briefly reviewed in terms of a universal  quasi-particle approach.
The case of a photon gas inside an appropriate black-body cavity is found to provide a physical interpretation of the lines spectra of atomic elements within an exact statistical approach based on a physical interpretation of the fine structure constant, $\alpha =1/137.0560$.  
\end{abstract}
%
\maketitle
\section{Introduction} 
Starting from the classical noise model of Nyquist (1928)~\cite{nyquist28} for a 3D conducting sample at
thermal equilibrium, we generalize the spectral density of current fluctuations to the case of quantum--relativistic conditions that avoid vacuum catastrophe associated with zero--point energy contributions at increasing frequencies of the electromagnetic spectrum.
Then, for the case of low frequencies, we generalize the electrical current--spectral density to the case of a 1D ballistic transport--regime that by recovering Landauer (1957) pioneer results~\cite{landauer57}
generalizes current fluctuations to the case of a set of non--interacting quasi-particles 
like: particle number (number $N$), neutral particle (mass $m$), charged particle (charge $e$), photons (energy $\epsilon$).
Finally we introduce noise and conductance for a black--body cubic cavity 
at thermal equilibrium.  By applying Planck statistics we propose an interesting scheme where the emission spectra of any atom can be interpreted as quantized shot--noise of the black-body cavity with a one--to--one  correspondence between the line energy
 and the thermal energy of the black-body.  Within this scheme the fine--structure constant $\alpha \approx 1/137$ is here found to be the inverse of the photon---number variance of the corresponding black--body cavity.
 
\section{The vacuum catastrophe and the Casimir effect}
From the classical Nyquist theorem published in 1928 for the spectral density of
current fluctuations we can write:
\begin{equation}
	S_I(f)=4 k_BT\mathrm{Re} [Y(f)]
\end{equation}
with $k_B$ the Boltzmann constant, $T$ the absolute temperature and $\mathrm{Re} [Y(f)]$ the real part of the admittance: i.e. the electrical conductance at the given frequency, $f$.
In 1951 this classical theorem has been extended to the quantum regime by Callen and Welton~\cite{callen51}, and successively by Kubo~\cite{kubo57,kubo66} in 1957. This extension takes the form:
\begin{equation}\label{eq:callen}
	S_I(f)=
	4 k_BT \mathrm{Re}[Y(f)]\left( \frac{x}{e^x - 1}  + \dfrac{1}{2} x\right)  = 
	S_I^P(f) + S_I^C(f)
\end{equation}
with $x=hf/(k_BT)$, $h$ being the Planck constant.
The second term in the second member of Eq.~({\ref{eq:callen}})  (i.e. $\frac{1}{2} x $) represents a vacuum--fluctuations (or zero--point) contribution implying a divergence of the noise  spectrum at increasing frequencies. This divergence (vacuum catastrophe), associated with an infinite number of photons, is analogous to the well-known ultraviolet catastrophe of the classical black--body radiation spectrum and is in contradiction  with experimental results and the Stefan--Boltzmann law.
The last member of Eq.~({\ref{eq:callen}}) expresses the spectral density as the sum of two contributions
due to Planck ($S_I^P(f)$) and Casimir ($S_I^C(f)$). 
Indeed the problem of the vacuum catastrophe can
be solved by introducing the quantum–relativistic Casimir effect, and justify
the omission of the zero-point contribution for macroscopic systems, thus recovering
the Planck distribution and the Nyquist relation in agreement with the experimental evidence.
We recall that the Casimir force is the attraction of two uncharged material bodies due to modification
of the zero-point energy associated with the electromagnetic modes in the space between them~\cite{casimir48,lamoreaux05}. 

In 2017 and 2020 Reggiani and Alfinito suggested that the formulation of
the fluctuation–dissipation theorem that best describes macroscopic phenomena is
the one in which the zero-point contribution is neglected and that the Casimir contribution is in general neglible under normal conditions of temperature and volume~\cite{reggiani17,reggiani20}.
%
In this we differ from the approach of Callen-Welton and Kubo who did not evaluate the finite value of $S_I^C(f)$ in the spirit of the Casimir effect.
As a consequence, $S_I^C(f)$  can be considered as a quantum correction to the Nyquist law that was originally obtained by using classical thermodynamics and classical electromagnetism only, thus without invoking explicitly any quantum arguments. 
The fact that for macroscopic systems $S_I^C(f)$ takes negligible values is a reason why till now its contribution was not explicitly detected from fluctuations spectra at thermal equilibrium.  However, for microscopic systems (i.e. in the case of  atomic scale lengths) and sufficiently low temperatures,  its effect should become relevant, as evidenced by experiments on the Casimir effects~\cite{milton01,edery06,schmidt08,auletta09}.
An experimental analysis of the cross-over from the  Planck to the Casimir contribution remains a mandatory issue to be investigated.
%
%
%
%
%
\section{Noise in 1D  under ballistic transport} 
Here we consider the limit of $T \rightarrow 0$ to allow quantum and relativistic conditions to set in with the temperature  remaining the source of thermal  equilibrium fluctuations.
Starting from the classical case, we recall that 
for 1D ballistic (b) quasi-particles (qp) at $f \rightarrow 0$ it is:
\begin{equation}
S^b_{Iqp}
=(\mathrm{qp})^2 \frac{4{\overline {\delta N^2}}}{\tau_t}
= 4k_BT  G_{qp}^b \label{eq:si}
\end{equation}
with $I_{qp}$ the qp--current fluctuations  measured in the outside short--circuit,  $\tau_t$ the transit--time between opposite contacts of the device of length $L$, that following the fluctuation--dissipation theorem, implies
\begin{equation}
G^b_{qp}=(\mathrm{qp})^2 \frac{\overline {\delta N^2} }{h_c}
\end{equation}
with 
\begin{equation}
h_c = m \sqrt {\overline{u_x^{'2}}} L
\end{equation}
the classical dynamic action,
with the  quasi-particles  mean-squared differential (with respect to carrier number) quadratic velocity component along the $x$ direction, $\overline{ u^{'2}_x}$, properly defined for the classical, quantum and relativistic cases \cite{gurevich79,reggiani16}.
\par
With the above formalism, we move to the 1D quantum case by taking the following limiting condition:
(i) $h_c=h$, that is the classical action goes to the Planck constant;
$\overline {\delta N^2} = \overline {N}=2$, as dictated by transverse quantum-size conditions and limiting to consider the first band only 
including spin degeneracy \cite{greiner00}.
Then,  $G_{qp}^b$ becomes the fundamental unit of the 1D quantum ballistic conductance of the quasi-particle gas
given by:
\begin{equation}
G_{qp}^b=(\mathrm{qp})^2 \frac{2}{h}
\end{equation}
\par
Accordingly, we can define universal units of quantum conductances by introducing different type of quasi--particles : number $N$, charge $e$, mass $m$, photon--energy $\epsilon$, electromagnetic vacuum. 
Table 1 summarizes  these universal units of quantum conductance for a 1D ballistic conductor of length $L$ for a  quantum gas of non-interacting  quasi-particle (qp) of single spin/polarization.
Being at thermal equilibrium, the associated spectral density of current fluctuations are obtained by multiplying the given conductance by $4k_BT$.
%
\begin{table}[h]
\caption{
Fundamental units of quantum conductance for a 1D ballistic conductor of length $L$ for a  quantum gas of non-interacting  quasi-particles (qp) of given spin/polarization: particle number (N), massive (m), electrical (e), single photon-energy ($\epsilon$),  with $\varepsilon_0$ the vacuum permittivity.
} 
\begin{center}
{\begin{tabular}{@{}cc@{}} \toprule 
	  qp &  Conductance $G_{qp}$  
		\\
		\colrule
	$N$ &  $\dfrac{1}{h} $ \\	\\	
	$e$ &  $\dfrac{e^2}{h} $ \\	 \\
 	$m$ &  $\dfrac{m^2}{h}$   \\ \\
  $\epsilon$ & $\dfrac{\epsilon^2}{h} $ \\	\\
	$\mathrm{em-vacuum}$	& $\varepsilon_0 c$  \\
	\botrule	
		\end{tabular}}
	\end{center}
\label{tab:ger}
\end{table}
\section{Relativity and the photon gas}

Within a quasi-particle approach, by analogy with the electrical case 
\cite{reggiani16} we
consider a black-body cubic cavity of side $L$ at thermal equilibrium. 
Using statistics and Planck distribution law it can be shown that the following relation between the average number of photons $\overline N$ and its variance $ \overline {\delta N^2}$ holds~\cite{leff2002,leff2015}:
\begin{equation}
\overline {\delta N^2}
= \gamma \overline{N}
= \frac{1}{3}\left(\frac{2 \pi k_B}{ch}  LT\right) ^3
\label{eqplanck1}
\end{equation}
with $\gamma \approx 1.37$ the Fano factor. 

In the following, we propose to correlate the continuum spectrum of a black--body of given side and temperature containing a gas of photons described by Planck's statistics to the line spectrum of a single atom traditionally associated with the fine structure constant $\alpha$ (where $1/\alpha\simeq 137$) as~\cite{kinoshita1996}:  

\begin{equation}
	\overline {\delta N^2}
	= \gamma \overline{N}
	= \frac{1}{3}\left(\frac{2 \pi k_B}{ch}  LT\right) ^3
	=\dfrac{1}{\alpha_N}
	\label{eqplanck2}
\end{equation}

with $\alpha_N$ the fine--structure constant for the given average photon number. Equation~(\ref{eq:si}) for the ballistic spectral density of photon  current  fluctuations, $S^b_{Iph}$, gives:

\begin{equation}
S^b_{Iph}
= 4 (hf)^2\frac{\overline {\delta N^2}}{\tau_t}
= 4 k_BT G^b_{ph}
= 4k_BT (hf)^2\frac{\gamma \overline {N}}{h}
\end{equation}
with $f$ the photon frequency, $\tau_t=L/c$ the photon transit time along $L$, $G^b_{ph}$ the 1D photon ballistic conductance.
Accordingly, for the photon number spectral density we obtain:
\begin{equation}
\tau_t S^b_{Nph}
=\overline {\delta N^2}
= \gamma \frac{\overline{N}{\tau_t}}{\tau_l}
= \frac{1}{3}\left(\frac{2 \pi k_B}{ch} LT\right) ^3
= \frac{1}{\alpha_N}
\label{eqplanck}
\end{equation}
where we have introduced the photon lifetime $\tau_l=
h/(k_BT)$. We remark that Eq.~(\ref{eqplanck}) implies two constraints : the former one
\begin{equation}
\frac{\tau_l}{\tau_l} = 1
\end{equation}
where $\tau_t$ and $\tau_l$ represents, respectively, a property of the cavity and a property of the contacts
(the walls of the black-body), and the latter one (for $1/\alpha_N=137$)
%
%
%
\begin{equation}\label{eq:lt}
L T = 1.7 \times 10^{-2} \ (\mathrm{m\ K})
\end{equation}
that is reminiscent  of the Wien law by taking $L=2 \pi \lambda$.
The physical implications of these constraints can be considered at present an open problem needing further investigation. 
Finally, for the spectral density of the photon number fluctuations we can  write the expression representing the fluctuation--dissipation relation for the black body:
\begin{equation}
\tau_t S^b_{Nph}
= \overline {\delta N^2}
= \gamma \overline {N}
\end{equation}
\section{Conclusions and remarks} 
Starting from the classical noise model of Nyquist  and a reformulation of Drude electrical conductance  for a 3D conducting sample at
thermal equilibrium~\cite{drude900}, we have generalized the spectral density of current fluctuations to the case of quantum-relativistic conditions that avoid the vacuum catastrophe associated with the zero-point energy contribution at increasing frequencies of the electromagnetic field by introducing the quantum--relativistic Casimir force.
\par
Then, for the case of low frequencies, we have generalized the electrical current spectral density and the electrical conductance to the case of a 1D ballistic transport-regime for
a set of non interacting quasi-particles, specifically:
dimensionless particle number $N$, neutral atomic and sub-atomic  particle mass $m$, charged quasi-particle with charge $e$, relativistic neutral quasi-particles like photons with energy $\epsilon$.
\par
Finally, we propose an interesting definition of photon conductance and photon noise for a black-body cubic cavity at thermal equilibrium with given volume and temperature that could help to interpret recent experiments on photon counting statistics~\cite{reulet16,reulet24} and able to provide a physical interpretation of the fine and hyperfine structure constants evidenced in the atomic line spectra of hydrogen and pertaining to any atomic element of the periodic Mendeleev table. Indeed, in our model the fine and hyperfine spectral constants coincide with the inverse values of the respective variance of photon number fluctuations. 
	\begin{figure}[h]
	\centering
	\includegraphics[width=1\textwidth]{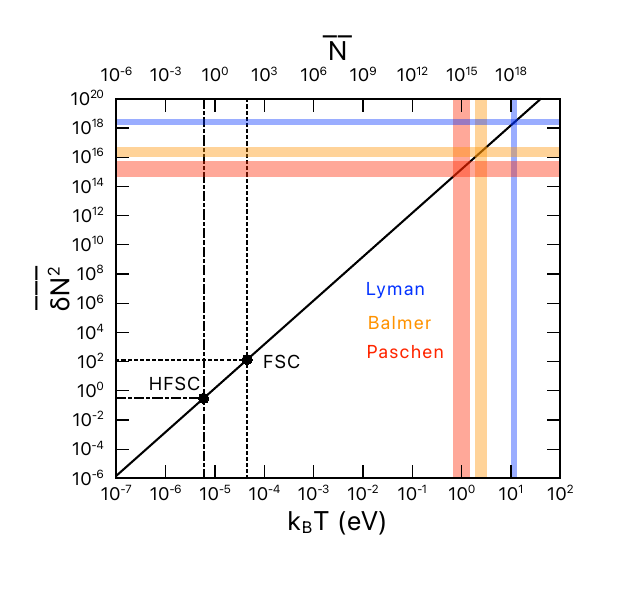}
	\caption{Plot summarizing  the variance of the photon numbers $\overline {\delta N^2}$ (or the inverse fine--structure constant of the H atom $1/\alpha$) as a function of the photon energy $k_BT$ and average photon number $\overline N$ including the available emission spectra of the Paschen, Balmer and Lyman series (colored shades areas),   obtained by  the statistical scheme property of the black--body  photon gas  as described by Planck's distribution law.  The universal behavior of the Fano factor is evidenced by the straight line representing the law: $\overline{\delta N^2} = \gamma \overline{N}$. 
		Analogous spectra of other atoms pertaining to the Mendeleev periodic table could be shown in the same way.	
		The two large points correspond to the values of the hydrogen fine-structure constant (FSC) and hyperfine--structure constant (HFSC). The reverse procedure should require the determination of the average number of photons under the conditions of temperature and volume $L_{\frac{1}{\alpha}} T_{\frac{1}{\alpha}} = 1.7 \times 10^{-2} \ (\mathrm{m\ K})$ of the black-body at thermal equilibrium.}
	\label{fig1}
\end{figure}	
The plot reported in Fig.~\ref{fig1} illustrates the graphical identification of the fit to obtain the value of $\overline{N}_{\frac{1}{\alpha}}$ that satisfies Eq.~(\ref{eqplanck}) within the same accuracy of the given value of $\frac{1}{\alpha}$.
Then, from Eq.~(\ref{eqplanck}) the proper value of the quantity $L_{\frac{1}{\alpha}} 
T_{\frac{1}{\alpha}}$ is obtained as presented, for instance, in Eq.~(\ref{eq:lt}). According to convenience, the single values of 
$L_{\frac{1}{\alpha}}$ and $T_{\frac{1}{\alpha}}$ can be obtained once their product is 
determined by the fitting.
As a significant example, by taking $\frac{1}{\alpha} = 137.0$ and $L_{\frac{1}{\alpha}} = 0.033 \ \mathrm{m}$ it is $T_{\frac{1}{\alpha}}=0.5156\  \mathrm{K}$ and $\overline{N}_{\frac{1}{\alpha}} = 100.1$.
\par
For completeness, the Table~\ref{tab:fununi} reports the numerical values used for the fundamental constants  and the corresponding physical quantities evaluated within the Planck's law for photon statistics used in Fig.~\ref{fig1}

\begin{table*}[t]
\caption{Relevant energy values of Hydrogen spectra and equivalent temperature together with the corresponding average photon numbers $\overline N$ used in Fig.~\ref{fig1}.
} 
\vskip 2pt
\begin{center}
{\begin{tabular}{@{}llll@{}} \toprule 
	    Spectral-type & Temperature (K) & Energy (eV)  &$\overline N $ 
		\\
		\colrule
	HFSC	& $6.9\times 10^{-2}$& $5.9\times 10^{-6}$ & $2.3\times 10^{-1}$\\
	FSC & $5.2\times 10^{-1}$ & $4.5\times 10^{-5}$& $1.0\times 10^{2}$\\
	Paschen& $0.77\div 1.8\times 10^{4}$ & $0.66\div 1.5\times 10^{0}$& $0.32\div 3.8\times 10^{15}$ \\
	Balmer & $2.2\div 3.7\times 10^{4}$ & $1.9\div 3.2\times 10^{0}$ & $0.74\div 3.7\times 10^{16}$\\
	Lyman & $1.2\div 1.6\times 10^{5}$ & $1.0\div 1.4\times 10^{1}$ & $1.2\div 2.8\times 10^{18}$
	\\
	\botrule	
		\end{tabular}}
	\end{center}
\label{tab:fununi}
\end{table*}

\section{Acknowledgments}
Prof. Tilmann Kuhn from M\"unster University is warmly thanked for the very valuable comments he provided on the  subject. 
\end{document}